\documentclass[aps,pra]{revtex4}
\usepackage{graphicx}

\newcommand{\tO}{\widetilde{\Omega}}
\begin{document}
\title{Multiply quantized vortices in trapped
Bose-Einstein condensates}
\author{Emil Lundh}
\affiliation{Department of Theoretical Physics,
Ume{\aa} University,
SE-901~87 Ume{\aa}}
\begin{abstract}
Vortex configurations in
rotating Bose-Einstein condensed gases trapped in power-law
and anharmonic potentials are studied. When the confining potential 
is steeper than harmonic in the plane
perpendicular to the axis of rotation, 
vortices with quantum numbers larger than 
one are energetically favorable if the interaction is weak 
enough. Features of the wave function for small and
intermediate rotation frequencies are investigated numerically.
\end{abstract}


\maketitle

\section{Introduction}
Trapped Bose-Einstein condensed gases provide a novel kind of
condensed matter system which can sustain quantized vortices, as
has been realized in the recent years \cite{jila,ens1,ens2,mit}.
Features which differ from those in macroscopic superfluid systems
are expected here. The fact that these systems are finite, and
that the parameter regime where the vortex core size is comparable
to the system size is attainable, in principle opens up for the
possibility for vortices in these systems to have a circulation
quantum number $q$ larger than unity.

It is well known that multiply quantized vortices are not
thermodynamically stable in spinless, macroscopic and homogeneous
superfluids, because the energy of a vortex
depends on the square of the circulation, and therefore two singly
quantized vortices with a finite spatial separation
have lower energy than one doubly quantized,
while giving the system the same angular momentum \cite{donnelly}.
However, multiply quantized vortices may well be energetically
favorable in systems which do not fulfill the criteria of
being homogeneous, spinless and macroscopic.
The assumption of homogeneity is not met in
superconductors with pinning forces, where indeed doubly
quantized vortices are observed \cite{scpinning}.
In $^3$He-A, where the order parameter is not a scalar due to spin
degrees of freedom, a lattice of doubly quantized vortices
with filled cores
has recently been observed \cite{helsinki}.
The argument also fails in spatially confined systems
where the vortex cores are not
much smaller than the system, such as mesoscopic
superconducting disks, where vortices with large quantum numbers
have been predicted \cite{schweigert}.

In condensed Bose gases confined in harmonic-oscillator potentials,
it has been found both analytically \cite{pethick},
numerically \cite{butts,castindum} and variationally \cite{castindum}
that multiply quantized vortices are not energetically favorable,
and no quantum numbers
larger than unity have been observed in experiment \cite{ens2,mit}.
However, none of these studies has considered whether altering the
power law of the confining potential can open up for the existence
of multiply quantized vortices.
In this paper, we shall give rigorous criteria for the
thermodynamic stability of multiply quantized vortices in trapped
Bose-Einstein condensates, and show the decisive role played by
the shape of the potential.

The system under study is a gas of bosons of mass $m$ in an
external potential $V(r)$, dilute enough and at
sufficiently low temperature that it is well described by the
Gross-Pitaevskii equation \cite{gross,pitaevskii,thebook}
\begin{equation}\label{gpe}
  i\hbar\frac{\partial \psi(r,t)}{\partial t} =
  -\frac{\hbar^2}{2m}\nabla^2\psi(r,t) + V(r)\psi(r,t) +
  U_0 |\psi(r,t)|^2 \psi(r,t)
  - \Omega\hat{L_z}\psi(r,t).
\end{equation}
The so-called condensate wave function $\psi(r,t)$ is normalized
to the number of particles $N$ in three dimensions, and to
the number of particles $\nu$ per unit length in two dimensions.
The coefficient $U_0$ in front of the nonlinear term is the
interaction strength, defined as $U_0= 4\pi\hbar^2a/m$, where
$a$ is the s-wave scattering length.
It turns out that the effective measure of interaction strength in two
dimensions is the product $4\pi\nu a$, which we shall denote by
$g$. In three dimensions the corresponding quantity depends on the
trapping potential.

A centrifugal term is present in the Gross-Pitaevskii equation,
corresponding to a rotation of the trap with the
frequency $\Omega$ about the $z$ axis. At certain critical values
$\Omega_{\text{c}q}$
of the rotation frequency, we expect there to be a discrete
transition between states of different circulation numbers, so
that when $\Omega_{\text{c}q+1} > \Omega > \Omega_{\text{c}q}$,
a state with total circulation
quantum number $q$ is the ground state \cite{donnelly}. The
critical frequencies are functions of the coupling strength $U_0$
\cite{lps}.

When the Gross-Pitaevskii equation is valid, the total energy of
the system is given by the mean-field energy functional
\begin{equation}
\label{gpenergy}
  E = \int dr \psi^*(r)\left(-\frac{\hbar^2}{2m}\nabla^2 + V(r)
  -\Omega \hat{L}_z\right)\psi(r) + \frac12 U_0|\psi(r)|^4.
\end{equation}

We shall in the following be concerned with the lowest-energy
solutions of the Gross-Pitaevskii equation (\ref{gpe}) for
a range of values of the driving frequency $\Omega$, interaction
strength $U_0$ and for different external potentials $V$.
The paper is organized as follows.
In Section \ref{AnalysisSec}, we present an analytic study of
systems contained in three- and two-dimensional power-law and
anharmonic traps. In Sec.\ \ref{NumSec}, we illustrate these 
findings numerically for the two-dimensional case
and study some features of the different states with
small and intermediate circulation quantum numbers.
Sec.\ \ref{ConclusionSec} provides a conclusion.

\section{Vortex configurations in power-law and anharmonic traps}
\label{AnalysisSec}

Consider a Bose gas trapped in a cylindrically symmetric power-law
potential of power $n$, which in cylindrical coordinates is
written
\begin{equation}
\label{trap}
  V(r,\phi,z) = \hbar\omega \left(\frac{r}{d_t}\right)^n +
  V_{\parallel}(z).
\end{equation}
The trap frequency $\omega$, which determines the strength
of the potential, defines a trap length $d_t =
\sqrt{\hbar/m\omega}$. The dependence of the potential on the $z$
coordinate is totally arbitrary as long as the problem remains
separable in the noninteracting limit, as is the case here.
We shall hereafter neglect to mention
the irrelevant $z$-dependent term and its associated degrees of
freedom.

In the analysis of this section, we shall consider
the weakly-interacting limit where $U_0$ is small,
because if a multiply
quantized vortex can ever be present, it will be so when the
interaction is weak. To see why, note that the vortex cores are
large for weak interactions and vice versa \cite{thebook,bp}.
Therefore, in the limit of large $U_0$, the vortices are much
smaller than the system size and the radius of curvature of the
density profile. The result of a homogeneous bulk system thus
applies, namely that a vortex array is the energetically favorable
configuration. Therefore, we expect multiply quantized vortices to
show up only in the realm of small (and possibly intermediate)
$U_0$.

Ref.\ \cite{pethick} showed why multiply quantized
vortices are never the minimum-energy state in a system which is
harmonically confined in the $x$-$y$ plane. We will here extend the
same analysis to the case when the
confinement is steeper than harmonic.
The argument is based on perturbative
treatment of the interaction term in the Gross-Pitaevskii equation
(\ref{gpe}). We therefore first study the noninteracting gas.
When $U_0=0$,
the Gross-Pitaevskii equation is identical to the one-particle
Schr{\"o}dinger equation, whose solutions are well known when the
radial confinement is harmonic. The eigenstates $\varphi_{n_rq}$
can be labeled by a radial and an axial quantum number $n_r$ and
$q$ (as well as a quantum number associated with the $z$ direction,
irrelevant for our considerations). The eigenenergies including the
centrifugal term are
\begin{equation}
E_{n_rq} = \hbar(\omega-\Omega)q + \hbar\omega(1+n_r).
\end{equation}
We shall concentrate on the states with no radial nodes,
$\varphi_{0q}$, because these have the lowest energy for a given
angular momentum $Nq\hbar$. These are in fact
vortices with quantum number $q$. When
the driving frequency $\Omega$ is less than the trap frequency
$\omega$, the $q=0$ state is the ground state; when
$\Omega=\omega$, all the states $\varphi_{0q}$ are degenerate.
When $\Omega>\omega$, $E$ decreases with increasing $q$, and there is
no ground state. Hence, in the noninteracting case all the critical 
frequencies $\Omega_{\text{c}q}$ are equal to the trap frequency 
$\omega$.

Following Ref.\ \cite{pethick}, we now consider perturbatively
the interaction energy. The perturbation for a given wave function
$\psi$ is, from Eq.\ (\ref{gpenergy}),
\begin{equation}
H' = \frac12 U_0 \int dr |\psi(r)|^4.
\end{equation}
Calculating this for general superpositions
$\psi(r) = \sum_q C_q \varphi_{0q}(r)$, we find that the
degeneracy at $\Omega=\omega$ is lifted, and the
critical frequencies $\Omega_{\text{c}q}$ split apart and assume
values less than $\omega$.
Fixing a total angular momentum $L_z=Nq\hbar$, it turns out that
wave functions which are superpositions of states of
different $q$ have less interaction energy than the pure
multiply-quantized vortex configurations $\varphi_{0q}$
for $q\ge 2$. Such
superpositions are vortex arrays, and these are the energetically
favorable states for any finite interaction strength.

We shall now show how the analysis of Ref.\ \cite{pethick} works
when the trapping power $n$ in Eq.\ (\ref{trap}) is larger than two.
In this case, the single-particle energy varies faster than linear
with the angular momentum.
This implies that in the noninteracting case,
pure multiply-quantized vortex states $\varphi_{0q}$
have lower energy than any superposition having the same angular
momentum. Hence, it takes a finite (albeit possibly very small)
interaction strength $U_0$ to overcome this difference. We stress
that in order to arrive at this result, we did not need to know
any details of the eigenfunctions of a general power-law
potential: the fact that the energy depends stronger than linearly
on the angular momentum is enough.
For a finite but
weak enough interaction, the multiply-quantized vortex states
are the minimum-energy configurations within their respective ranges
of external rotation frequency $\Omega$. When the interaction is
strong enough, however, the larger interaction energy of the
multiply-quantized vortex states will overcome the difference in
kinetic plus trap energy, and the ground state will again be a vortex
array. This makes sense, since as we noted above, in the limit of
strong interaction the results of the infinite and homogeneous case
apply.

For potentials weaker than harmonic, $n<2$, no rotation is
possible. The potential in a frame rotating with angular frequency
$\Omega$ is $V(r) - \frac12 \Omega r^2$, which is not confining if
$\Omega>0$ and $V(r)$ is weaker than harmonic. Therefore a rotating
state in such a trap can be at most metastable. \cite{privatePethick}

If a pure power-law potential which is steeper than harmonic may 
seem an artificial construction which is hard to fabricate in 
practice, we note that the analysis can be repeated for an
anharmonic potential,
\begin{equation}\label{anharmpot}
V(r,\phi,z) = \frac12m\omega^2 r^2(1+\lambda \frac{r^2}{d_t^2})
+ V_{\parallel}(z).
\end{equation}
The important point is that for any positive
anharmonicity $\lambda$, the single-particle energy varies faster than
linear with the angular momentum. This again leads to the
conclusion that multiply quantized vortices can exist in anharmonic
traps as long as $\lambda>0$ and the coupling is small.

We have shown that multiply quantized vortices can be stable only in
trapping potentials which are steeper than harmonic in the $x$-$y$
plane, and only for weak enough interactions. A close inspection of
the above argument reveals the physical mechanism behind the
transition. The kinetic energy gives rise to a repulsion between the
vortices because of the overlapping velocity fields, while the
trapping potential strives to keep the vortices together in order to
reduce the spatial extent of the cloud. The interaction energy favors
states with several separated vortices because such states have lower
density on the average. The thermodynamic stability of multiply
quantized vortices is thus a consequence of the competition between
these three energies, and the power 2 is the limiting value.

\section{Numerical results for two-dimensional systems}
\label{NumSec}

We now perform a numerical study to quantify the predictions of
Section \ref{AnalysisSec}. We shall treat the two-dimensional case
only. We insert the power-law potential, Eq.\
(\ref{trap}), into the Gross-Pitaevskii equation (\ref{gpe}) and
scale out the dimensions by defining $r=d_t \tilde{r}$,
$t=\omega^{-1}\tilde{t}$,
$\psi = \sqrt{\nu}d_t^{-1}\tilde{\psi}$, $L_z=\hbar\tilde{L}_z$,
$\Omega=\omega\tO$, $g=4\pi \nu a$, and for the total mean-field
energy $E = \hbar\omega \tilde{E}$.
There results a dimensionless Gross-Pitaevskii equation,
\begin{equation}\label{dimlessgpe}
  i\frac{\partial \tilde{\psi}}{\partial \tilde{t}} =
  -\frac{1}{2}\nabla_{\tilde{r}}^2\tilde{\psi} +
  \frac12\tilde{r}^n\tilde{\psi} +
  g |\tilde{\psi}|^2 \tilde{\psi}
  - \tO\hat{\tilde{L}_z}\tilde{\psi},
\end{equation}
which depends on three parameters, namely the coupling $g$,
the rotation frequency $\tO$, and the trapping power $n$. We
solve it in two dimensions using the split Fourier method
\cite{castindum}. By propagating Eq.\ (\ref{dimlessgpe})
in complex time, the system relaxes to a
local energy minimum which depends on the initial state. In order
to find the {\em global} minimum-energy configuration for each
choice of parameters $g$, $\tO$, $n$, we perform the relaxation
a few times using different, irregular initial states, compute
their energy with the aid of Eq.\ (\ref{gpenergy}), and out of
the different final states we choose that which has the lowest
energy. The grid size is $64\times 64$ points, and the lattice
constant is equal to 0.2 times the trap length $d_t$. We have
confirmed that doubling the grid size and halving the lattice
constant does not alter the results.

\begin{figure}
\includegraphics[width=\columnwidth]{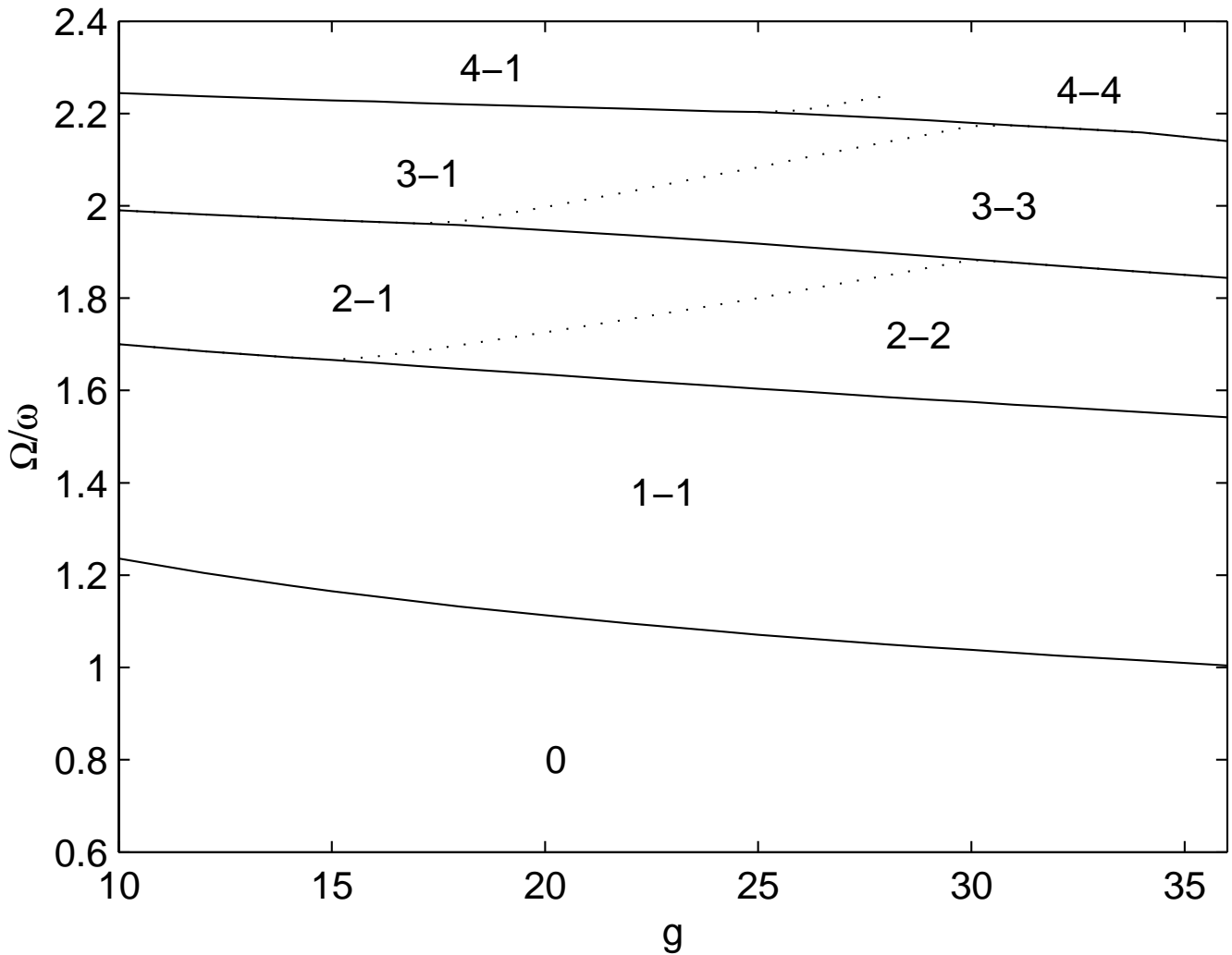}
\caption{\label{phdiag4}Phase diagram for a quartically confined
two-dimensional Bose gas subject to a force rotating with an
angular frequency $\Omega$. The state marked by 0 is the
nonrotating state, and 1-1 is a state with one central $q=1$
vortex. The regions 2-2, 3-3 and 4-4 denote
vortex lattices containing two, three and four singly quantized
vortices, respectively, while the states 2-1, 3-1 and 4-1 are
states with several quanta of circulation but only one phase
singularity.}
\end{figure}

Figure \ref{phdiag4} is a phase diagram for the quartic potential
($n=4$), which shows how the $\tO$-$g$ plane is divided into
regions of different vortex configurations with the total
circulation $q$ ranging from 1 up to 4. The different vortex
configurations are labelled by the total number of quanta
followed by the number of singularities present in the system: the
configuration 3-3 thus denotes three $q=1$ vortices whereas 3-1 is
a state with one triply quantized vortex present. The region
marked by 0 is the nonrotating state.
Since the numerical computation is done with limited precision,
we need an operational criterion for the existence of a multiply
quantized vortex. A $q$-fold quantized vortex is defined as a
configuration in which the calculated circulation around a circle
enclosing the origin is equal to $q$ and only one density
minimum can be seen on optical inspection of the plotted wave
function, i.\ e.\ the inter-vortex distance is less
than the lattice constant of the numerical grid.
The transition frequencies found using
this criterion coincides with the point where the angular
momentum saturates at an integer value.

\begin{figure}
\includegraphics[width=\columnwidth]{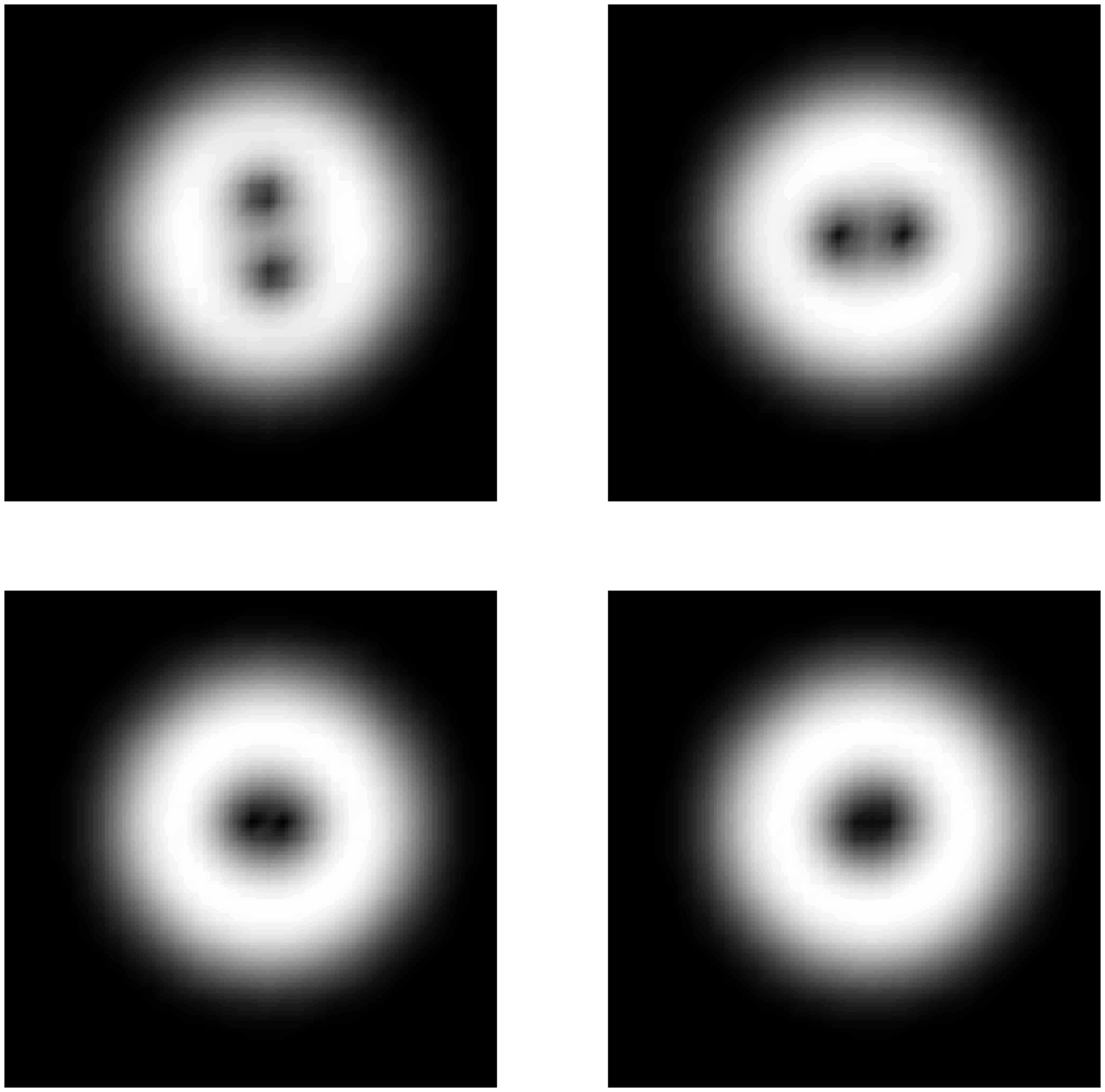}
\caption{\label{merge2}Numerically computed condensate wave
functions, illustrating how the
distance between two vortices continuously decrease as the
external rotation frequency $\Omega$ increases. The bright shades
indicate high density and vice versa.
The confinement is quartic, the coupling constant
$g=28.0$, and $\Omega/\omega = 1.59$ for the top left panel, 1.72 (top
right), 1.83 (bottom left) and 1.86 (bottom right). The latter is
the splitting frequency, operationally defined as the frequency
where no mode of plotting the wave
function reveals more than one density minimum.}
\end{figure}

As expected from the analysis of the previous section,
we find multiply quantized vortices for small
values of $g$, which split into arrays of singly quantized
vortices for larger $g$. The transition from a multiply quantized
vortex to an array is continuous, as illustrated in Figure
\ref{merge2}, where two vortices are seen to merge as the critical
line is crossed.

\begin{figure}[htbp]
\includegraphics[width=\columnwidth]{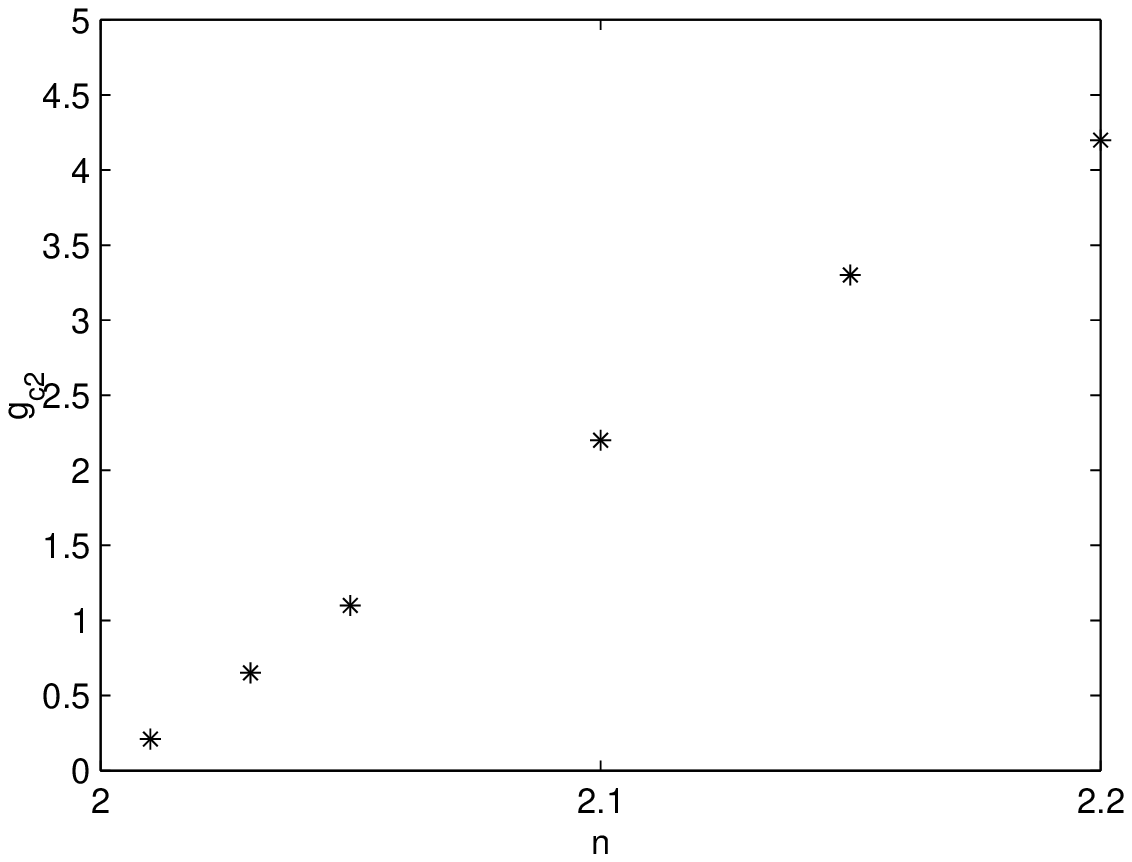}
\caption{\label{critover}Critical coupling for the stability of
doubly quantized vortices, $g_{\text{c}2}$, as a function of the
power $n$ for a condensate in a power-law trap. The analysis
of Sec.\ \ref{AnalysisSec}
predicts that the curve goes to zero when the trapping power
approaches 2, in accordance with these numerical results.}
\end{figure}

It is convenient to define a new quantity to describe
the regime of stability of multiply quantized vortices.
The $q$-th critical coupling $g_{\text{c}q}$ is defined so that when
$g>g_{\text{c}q}$, there can exist no $q$-fold quantized vortices
for any value of $\Omega$,
but when $g<g_{\text{c}q}$, a $q$-fold quantized vortex is the energy
minimum within certain limits of $\Omega$. The critical coupling
is a function of the trap power $n$.
Fig.\ \ref{phdiag4} suggests that the critical coupling increases
with $q$. The physical reason for this is, that if the system is to
accommodate many singly quantized vortices, the cores must be small
and $g$ large. Our analysis in the
previous section showed that $g_{\text{c}q}=0$ for the harmonic
trap, $n=2$, and $g_{\text{c}q}>0$ when $n>2$. This is confirmed in
Figure \ref{critover}, which shows how $g_{\text{c}2}$ depends on $n$.
The data points suggest that the curve $g_{\text{c}2}(n)$ goes
to zero linearly when $n \rightarrow 2$ from above. 

\begin{figure}
\includegraphics[width=\columnwidth]{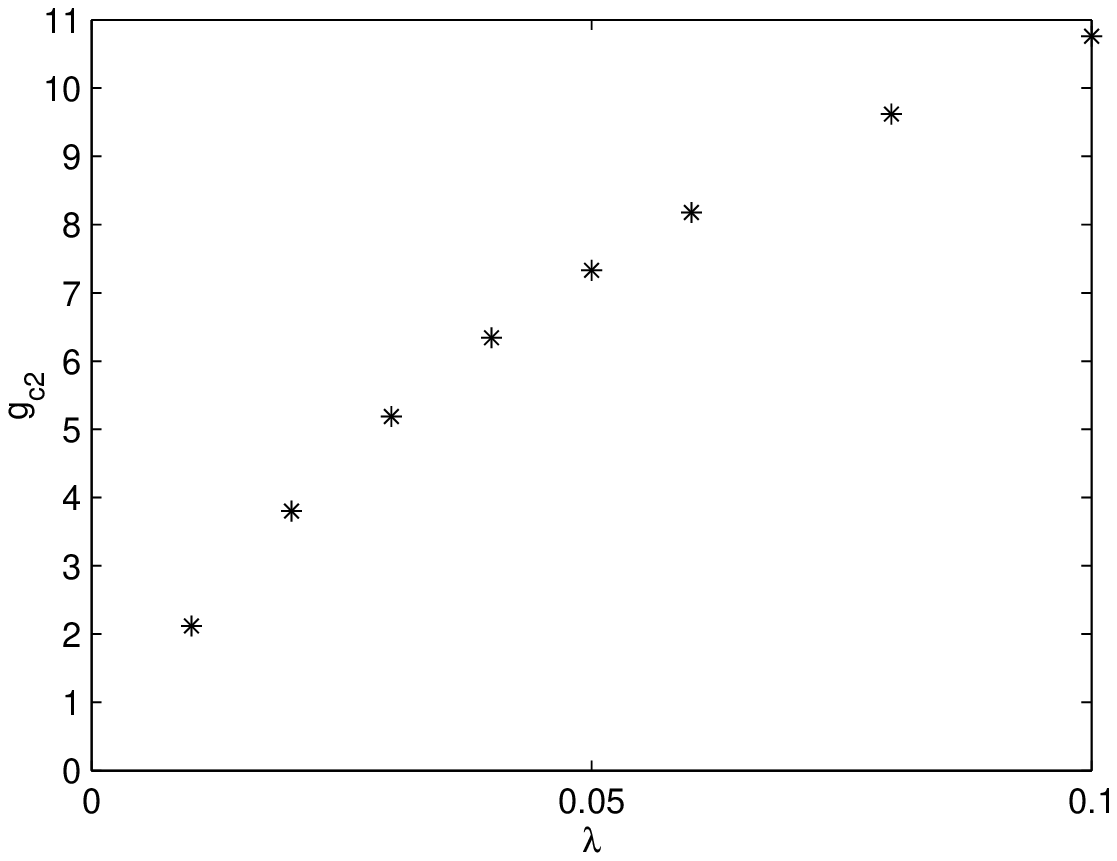}
\caption{\label{critan}Critical coupling $g_{\text{c}2}$ for a
condensate trapped in an anharmonic potential as a function of the
coefficient $\lambda$ of the quartic term.
When $\lambda$ approaches zero, the trap approaches the harmonic
case, and the critical coupling drops to zero.}
\end{figure}

We have pursued the analogous anharmonic case, Eq.\
(\ref{anharmpot}), using the same numerical procedure.
The qualitative features of the phase diagram are not different
from the pure power-law case. The critical coupling
$g_{\text{c}q}$ is now a function of the anharmonicity $\lambda$.
The result for $g_{\text{c}2}$ is shown in Figure \ref{critan}.
The limit $\lambda\rightarrow \infty$ is the quartic potential,
whose critical coupling $g_{\text{c}2}=30.5$ can be read off from
Fig.\ \ref{phdiag4}.
When $\lambda\rightarrow 0$, on the other hand, the potential
approaches a harmonic one, and $g_{\text{c}2}$ approaches zero.
In the experimentally realized  two-dimensional condensates of 
Ref.\ \cite{mit2d}, $g$ varies between 100 and $10^4$, the former for a
number of particles $N=10^4$. Diminishing
the number of particles by a factor 10 and introducing an
anharmonicity larger than 0.1 would take these systems 
below $g_{\text{c}2}$.

\begin{figure}
\includegraphics[width=\columnwidth]{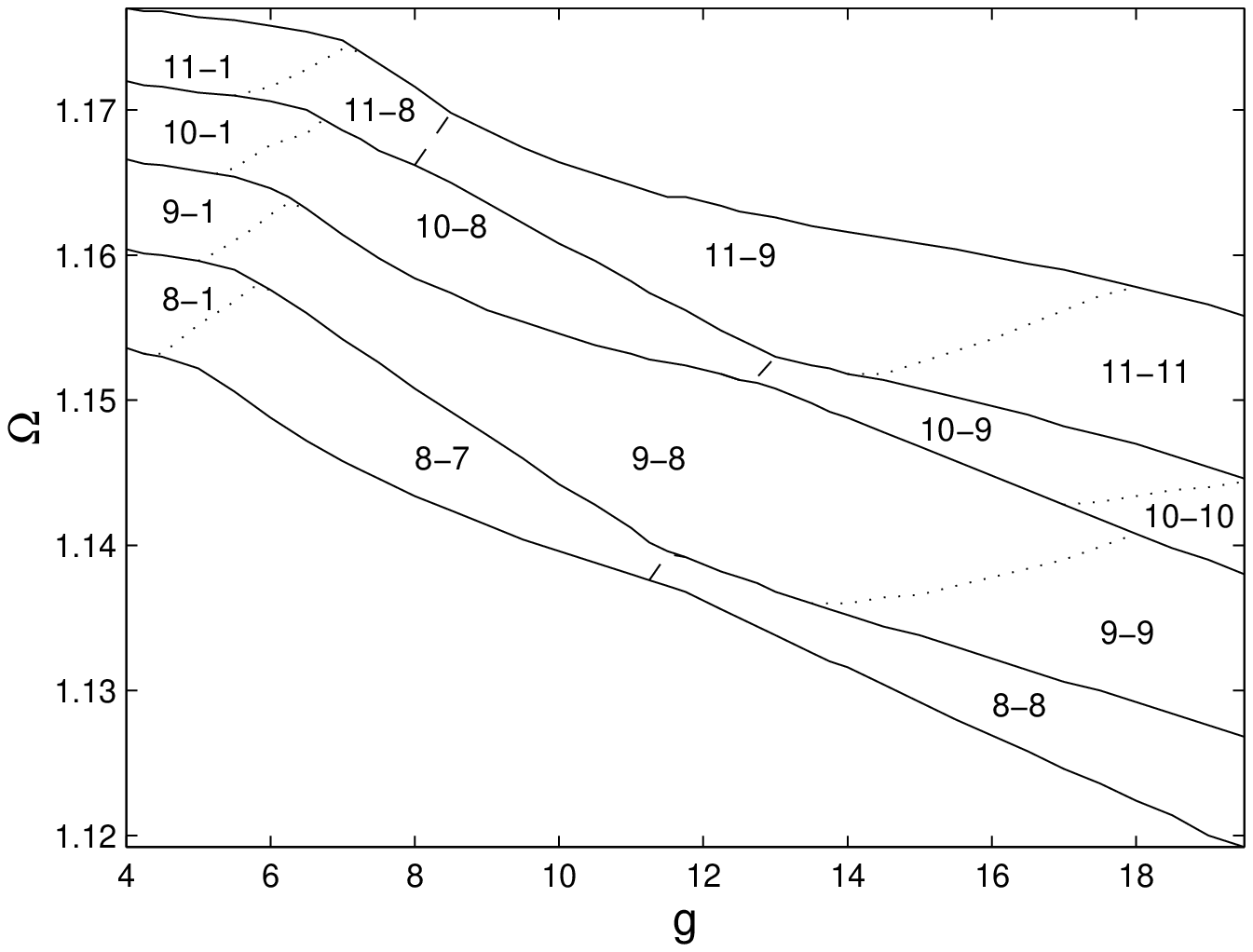}
\caption{\label{phdiag8}Phase diagram for a rotated condensate in
a power-law trap of power $n=2.2$. Full lines
indicate the critical frequencies separating states of different total
circulation. Dashed lines mark discrete transitions between states of
different symmetry but equal total circulation. Dotted lines mark
continuous transitions between states of equal total circulation but
different number of singularities. Only the critical lines for
circulation $q$ between 8 and 11 are shown. The different phases
are labelled by their total circulation, followed by the number of
singularities in the system; phases such as 8-7 thus contain both
singly and multiply quantized vortices in regular patterns.}
\end{figure}

The phase diagram in $g$-$\Omega$ space becomes richer
for higher quantum numbers $q$. For $q\le 7$, there
are only two phases, a vortex-array phase and a
multiply-quantized vortex phase. For higher quantum
numbers, however, a new kind of pattern appears,
where one multiply quantized vortex is surrounded by
singly quantized vortices. Figure \ref{phdiag8} shows the
different vortex phases for $8 \le q \le 11$.
The power of the potential is chosen to $n=2.2$.
We find a number of distinct phases, which we label by their total
circulation followed by the number of singularities just as in
Fig.\ \ref{phdiag4}.
There are three phases containing eight quanta of circulation:
a central $q=8$ vortex for small $g$ (denoted 8-1);
an array of eight vortices for large $g$ (8-8);
and for intermediate $g$, a central $q=2$ vortex
surrounded by six $q=1$ vortices (8-7).
The transition from 8-1 to 8-7 is continuous: the six $q=1$ vortices
separate from the central one in a smooth way as the critical line is
crossed. The two phases 8-7 and 8-8, on the other hand, have different
symmetries and the transition between these two is discontinuous.
For the case of nine quanta,
there exist the corresponding phases 9-1 and 9-9, and
one intermediate phase 9-8 with a central $q=2$ vortex
and seven $q=1$ vortices. A 9-7 configuration with a $q=3$ vortex
surrounded by six $q=1$ vortices seems to be stable in a very
narrow region of width $\Delta g \sim 0.2$,
between 9-1 and 9-8, but with the present numerical
precision we have not been able to accurately determine the 
boundaries of this phase and it is left out of Fig.\ \ref{phdiag8}.
For the case q=10, on the other hand, one can clearly
distinguish four different configurations, and likewise for $q=11$.

\begin{figure*}
\includegraphics[width=\textwidth]{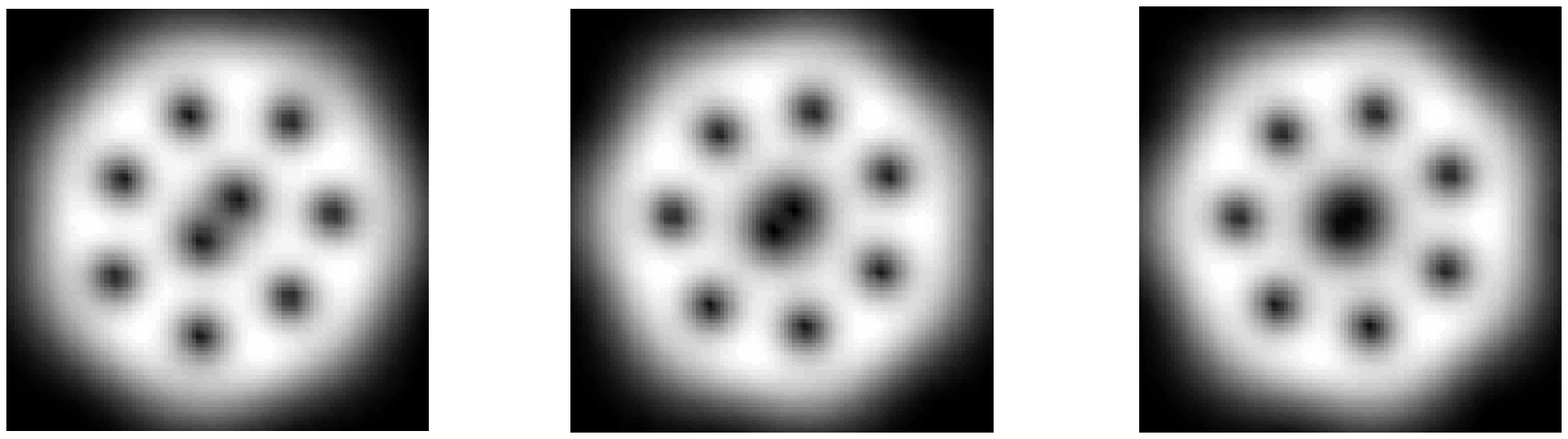}
\caption{\label{merge9}Density plots for a two-dimensional
condensate in a power-law trap of power 2.2, at coupling $g=19.5$.
The leftmost panel is the minimum-energy configuration for
rotation frequency $\Omega=1.1300\omega$, the middle panel shows
$\Omega=1.1400\omega$ and the right panel has
$\Omega=1.1460\omega$. All three states have a total circulation
$q=9$. The pictures show that the transition between the phases
9-9 (left and middle panels) and 9-8 (right panel) is continuous.}
\end{figure*}

Analogously to the $q=8$ case, the transition between 9-1 and 9-7
is continuous, and so is the transition between 10-1 and 10-8 and
between 11-1 and 11-8.
On the other hand, the configurations 9-7, 10-8 and 11-8 have
entirely different symmetries from 9-8, 10-9 and 11-9, respectively,
and those transitions are discontinuous. Finally, the three
last-mentioned states transform smoothly into arrays of singly
quantized vortices (9-9, 10-10 and 11-11) when $g$ increases, by
splitting of the central vortex into singly quantized ones. Figure
\ref{merge9} illustrates this.

\section{Conclusions}
\label{ConclusionSec}
We have determined the criteria for existence of vortices with
circulation quantum numbers larger than unity in rotated
trapped Bose-Einstein
condensates. We have proven that in both two- and three-dimensional
systems, multiply quantized vortices are the
energy minimum if the trapping potential in the plane perpendicular to
the axis of rotation is steeper than harmonic, and the
interaction is sufficiently weak so that the vortex cores are not much
smaller than the cloud, and the external rotation frequency is within
appropriate limits. For stronger interactions, the multiply quantized
vortices break up into arrays of several vortices. The physics of the
transition is the interplay between the inter-vortex repulsion and the
trapping potential. In the case of two-dimensional systems, we have
numerically determined the regimes of thermodynamic stability of
different vortex configurations. For clouds rotated at large angular
velocities, there are regimes where arrays containing both
singly and multiply quantized vortices are stable. 

Most trapping potentials used in experiments are harmonic, but
using Laguerre-Gaussian beams one can construct optical traps 
of arbitrary even power laws,
where our predictions can be tested \cite{kuga}. A two-dimensional
Sodium condensate can host doubly quantized vortices if the number of
particles is of the order 1000 or smaller, and a trap anharmonicity
of about 10\% is introduced. Alternatively, the appropriate 
weak-coupling regime can be attained by using Feshbach resonances 
to diminish the scattering
length \cite{feshbach}. In a three-dimensional system, a
quantitative determinition of the conditions has been prohibited by
numerical limitations.
The technique of vortex imaging by interference \cite{interference},
which directly images the phase of the condensate, promises to be a
convenient way to detect multiply quantized vortices.
Alternatively, one can look for multiply quantized vortices
by measuring the angular momentum of the system \cite{angmom}, or
simply by looking for anomalously large vortex cores.

\begin{acknowledgments}
I am grateful to my supervisor J{\o}rgen Rammer for supervising
this work, and Ping Ao for supervision at the early stages. I would
also like to thank Georgios Kavoulakis, C.\ J.\ Pethick, Thomas
Busch and Lars Melwyn Jensen for valuable discussions, NORDITA for
travel support and lodging, and Jani Martikainen for help with the
numerics.
\end{acknowledgments}

\end{document}